\documentclass{eas}
\usepackage{graphicx}

\begin{document}

\title{Metal enrichment in galactic winds}

\address{CEA Saclay, DSM/Dapnia/SAp, B\^atiment 709, 91191 Gif-sur-Yvette Cedex, France}
\author{Yohan Dubois$^1$}
\author{Romain Teyssier$^1$}

\runningtitle{Galactic Winds}

\begin{abstract} 
Observations give evidences of the presence of metals in the intergalactic medium (IGM). The stars responsible for transforming hydrogen and helium into more complex atoms do not form outside the galaxies in the standard scenario of galaxy formation. Supernovae--driven   winds  and   their
associated  feedback was  proposed as  a possible
solution to explain such enrichment of the IGM.   It  turned  out  that  a proper  modelling  of  supernovae
explosions within a turbulent interstellar medium (ISM) is a difficult
task.  Recent advances have  been obtained using a multiphase approach
to  solve for  the  thermal state  of  the ISM,  plus some  additional
recipes  to  account for  the  kinetic  effect  of supernovae  on  the
galactic  gas.   We briefly describe  here  our  implementation of  supernovae
feedback within  the RAMSES  code, and apply  it to the  formation and
evolution of  isolated galaxies of various masses  and  angular
momenta.  We have  explored under what conditions a  galactic wind can
develop, if one considers only  a quiescent mode of star formation. We have also characterized the distribution and evolution of metallicity in the gas outflow spreading in the IGM. 
\end{abstract}

\maketitle

\section{Introduction}

Supernovae--driven  winds  are  a  key ingredient  of  current  galaxy
formation  models, in  order to  suppress the  formation  of low--mass
galaxies and maybe to  solve the so--called ``overcooling'' problem in
the  current  hierarchical  scenario  of structure  formation  in  the
universe. The  proper modelling of  galactic winds is a  difficult and
unavoidable task, both in  semi--analytical models (Hatton {\it et  al.}  2003) and in
numerical simulation  (Springel \& Hernquist 2003,  Rasera \& Teyssier
2006) of  galaxy  formation.    Observational  evidence  for  galactic
outflows have already been pointed  out by several authors (Heckman et
al 2000, Adelberger {\it et  al.}  2003).  They are usually associated
to massive  starbursts, for which  very strong outflows  are reported:
for one solar mass of star formed in the galaxy, between 1 and 5 solar
masses  of  gas  are  ejected  in those  winds  (Martin  1999).   This
translates  into a  wind efficiency,  usually noted  $\eta_w$, ranging
from  100\% to 500\%.   The effect  of galactic  outflows can  also be
measured  in  the enrichment  of  the  intergalactic  medium (IGM)  as
observed in absorption lines of quasars spectra (Bouch\'e {\it et al.}
2006).

Despite the  difficulties of modelling supernovae  explosions within a
turbulent, multiphase and magnetized ISM, understanding the physics of
the resulting large scale outflows is also a challenge. Many questions
arise:  what are the  conditions for  a galactic  wind to  develop and
escape from the galaxy potential well~? What is the mass ejection rate
of  such a  wind~?  What  is  the metallicity  of the  wind and  other
associated observational  signatures~?  As explored by  Fujita {\it et
al.}  (2004) in the context  of an isolated, pre-formed galactic disc,
the  ram-pressure of  infalling material  might be  the  main limiting
factor for galactic winds to exist.

 Our  goal is  here to present  new simulations  of galactic
winds performed  with the RAMSES code, using  Adaptive Mesh Refinement
with a state-of-the-art shock-capturing scheme (Teyssier 2002).

In this paper, we follow the approach of Springel {\it et al.} (2003),
considering  an  isolated  Navarro,  Frenk  \& White  (1996)  halo  in
hydrostatic equilibrium, that self-consistently cools down and forms a
centrifugally supported disc  in its center.  Star  formation and the
associated supernovae explosions  proceed according to rather standard
recipies. 

\section{Numerical methods}
\label{nummeth}

We have considered that initially  gas and dark matter follow the same
NFW mass density profile, with 15\%  of the total mass in baryons in hydrostatic
equilibrium. The  concentration  parameter   was  set   to  $c=10$,
independant  of  the halo  mass. The  initial  angular momentum  is given by the fitting formula $j(r)= j_{max} M(r)/M_{vir}$ of Bullock et al.   (2001), normalized  to the  halo  spin  parameter $\lambda=J  \vert
E\vert ^{1/2}/GM^{5/2}_{vir}$, for which we have considered two cases:
$\lambda =0.04$ and $\lambda =0.1$.

The gas  is radiatively  cooled using a  metallicity-dependant cooling
function  (Courty \& Teyssier,  in prep.),  since metal  enrichment by
supernovae feedback  is self-consistently simulated in  this study.
 In order  to  take  into  account  the  thermal  feedback  of  supernovae
explosions, we  have implemented a  simplified form of  the multiphase
model of Springel \& Hernquist  (2003), using a polytropic equation of
state in the regions of star formation ($n_0 > 0.1$ cm$^{-3}$) and where the temperature is fixed   to  the   polytropic  equilibrium temperature.
In these regions a fraction of gas is converted into star particles  using a Schmidt law, with a time  scale parametrized by $t_*=t_0 ( n_H  / n_0 )^{-1/2}$. We chose $t_0=3 \, \rm Gyr$ or $8  \,  \rm  Gyr$ in our simulations,  resulting  in  different  overall  star  formation efficiencies (see Rasera \& Teyssier for details).

Kinetic feedback due to supernovae explosions that we have implemented in RAMSES is the key process of this study.
  For each  star particle formed, we create ``debris
particles'',  containing both  supernovae ejecta  and  surrounding gas
entrained by the blast wave.  The velocity of the ``debris particles''
is  computed according to  a local  Sedov solution.   Debris propagate
freely as  collisionless particles over a distance  corresponding to a
blast wave  radius of  2 cells. At  this moment, debris  release their
mass, metal  content, momentum  and energy to  the gas cell  they have
reached. 

We  have  performed isolated  galaxy  simulations  with two  different
virial  masses  $10^{10}  \,   \rm  M_{\odot}$  and  $10^{11}  \,  \rm
M_{\odot}$ in a box of 6 $R_{vir}$ and over 6 Gyr. The coarse grid of those simulations has $128^3$ cells and the maximum resolution is about 150 pc.

\section{A toy model for galactic winds}
\label{anaprb}

\begin{figure}[h]
  \includegraphics[width=6cm]{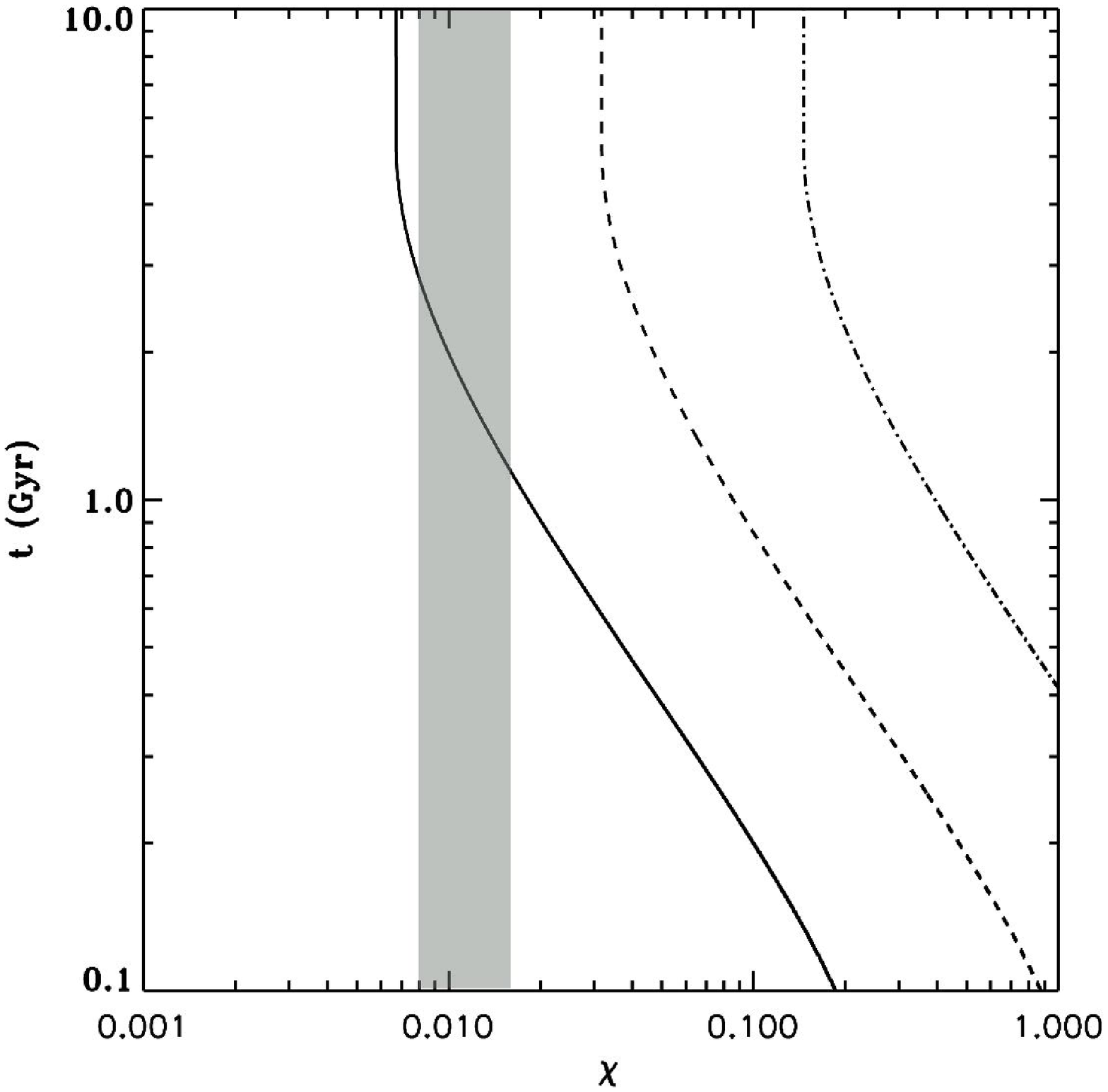}
  \includegraphics[width=6.5cm]{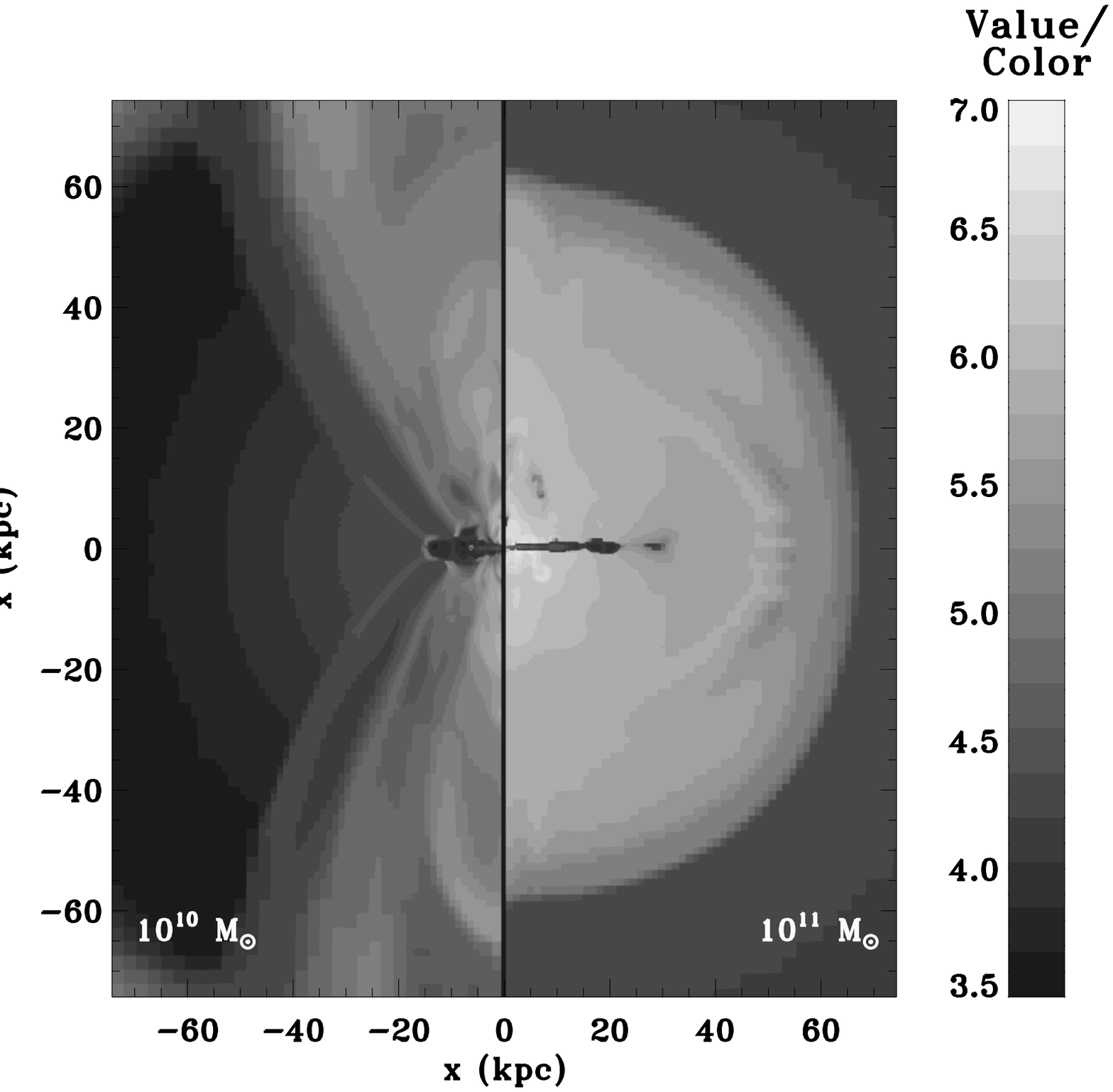}
  \caption{{\it  Left}:   Wind  break-out  epoch  as   a  function  of
  hydrodynamical  efficiency $\chi$ for  three different  halo masses:
  $10^{10} \, \rm M_{\odot}$,  $10^{11} \, \rm M_{\odot}$, $10^{12} \,
  \rm M_{\odot}$ (from left to  right). The shaded area corresponds to
  the   allowed    range,   as   demonstrated    by   our   numerical
  simulations. {\it Right}: Comparison of the temperature map obtained
  at $t=$  6~Gyr for 2  different simulations, with a  galactic
  wind  breaking out of  the low  mass halo (left),  and a  galactic fountain
  remaining close to the disc in the high mass case (right).}
  \label{figure_1}
\end{figure}
   
Following Fujita {\it  et al.}  (2004), we anticipate  that the infall
of gas  from the halo will  be the key  process in the success  or the
failure of galactic outflows.   Gas infall fuels the  star forming  disc with  fresh  gas, and  therefore controls  the
supernovae explosion  rate in the  galaxy but it is also the
source of ram  pressure that can confine any  outflowing material from
the disc. We  propose here a very simple model in  which the total gas
mass evolves according to the following ``open box solution''
\begin{equation}
M_g(t)=\int_0^t \exp(\frac{u-t}{t_*})\dot M_{acc}(u) {\rm d}u\, ,
~~~{\rm and}~~~\dot M_*(t) = \frac{M_g(t)}{t_*}\, .
\end{equation}
The accretion rate $\dot M_{acc}$ is computed assuming that each shell
of our  NFW halo  free-falls down  to the center,  where the  disc has
formed. We then compare the supernovae luminosity in the disc to the
accretion luminosity of infalling gas
\begin{equation}
L_w(t)=\chi \dot M_*(t) \eta_{sn} {E_{SN} \over M_{SN}} \, ,
~~~{\rm versus}~~~L_{acc}(t) = {1\over 2}\dot M_{acc} v_\infty^2\, .
\end{equation}
where $v_\infty$ is the  terminal velocity of free-falling gas shells,
computed using the  NFW mass profile. The key  parameter of the model,
$\chi$, is  called the ``hydrodynamical efficiency'' and  refers to the
conversion efficiency  of supernovae energy into  galactic wind energy.
$\eta_{SN}\simeq$ 10\% is the mass fraction of stars that explode into
type  II  supernovae,  according   to  standard  stellar  initial  mass
function, $E_{SN}\simeq 10^{51}$ erg is the typical energy produced by
one single  supernova, and $M_{SN}\simeq  10$ M$\odot$ is  its typical
progenitor mass.
A part of the energy of supernovae explosions creates  heating and  turbulence in the multiphase ISM. Since this energy dissipated in the ISM does not contribute in the global outflow, the hydrodynamical efficiency is likely to be small.
The fraction of energy that manage to
escape  from  the   dense  gaseous  disc  will  depend   on  the  disc
characteristic  (thickness, size, gas  content).  In  our simulations,
these properties  will be specified by  the the spin  parameter of the
halo and  by the  star formation  time scale of  the Schmidt  law.  In
Figure~\ref{figure_1}, the  wind break-out  epoch (i.e.  when  the wind
luminosity exceeds the accretion luminosity) is shown for various halo
masses, as  a function of  the unknown parameter $\chi$.   One clearly
sees that the smaller the hydrodynamical efficiency, the later the wind
will blow out  of the disc.  More importantely, for  each halo mass, a
minimum efficiency  is required  in order for  a wind to  appear: 0.7\%
(resp. 3\% and 15\%) for  a $10^{10}$ M$_\odot$ halo (resp.  $10^{11}$
and  $10^{12}$).

\section{Simulation results}
\label{results}
 
\begin{figure}[h]
  \centering
  \includegraphics[width=6cm]{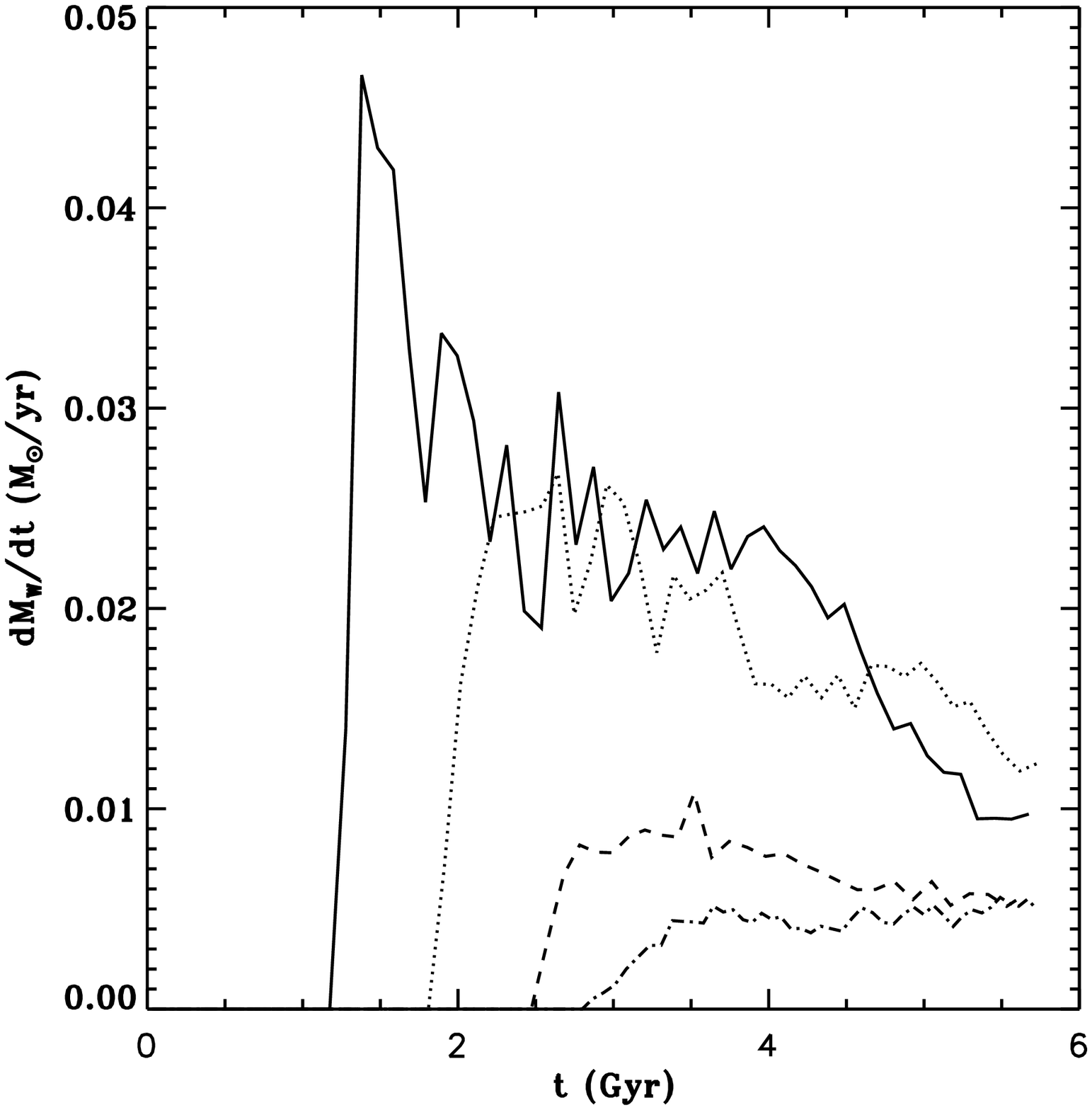}
  \centering
  \includegraphics[width=6cm]{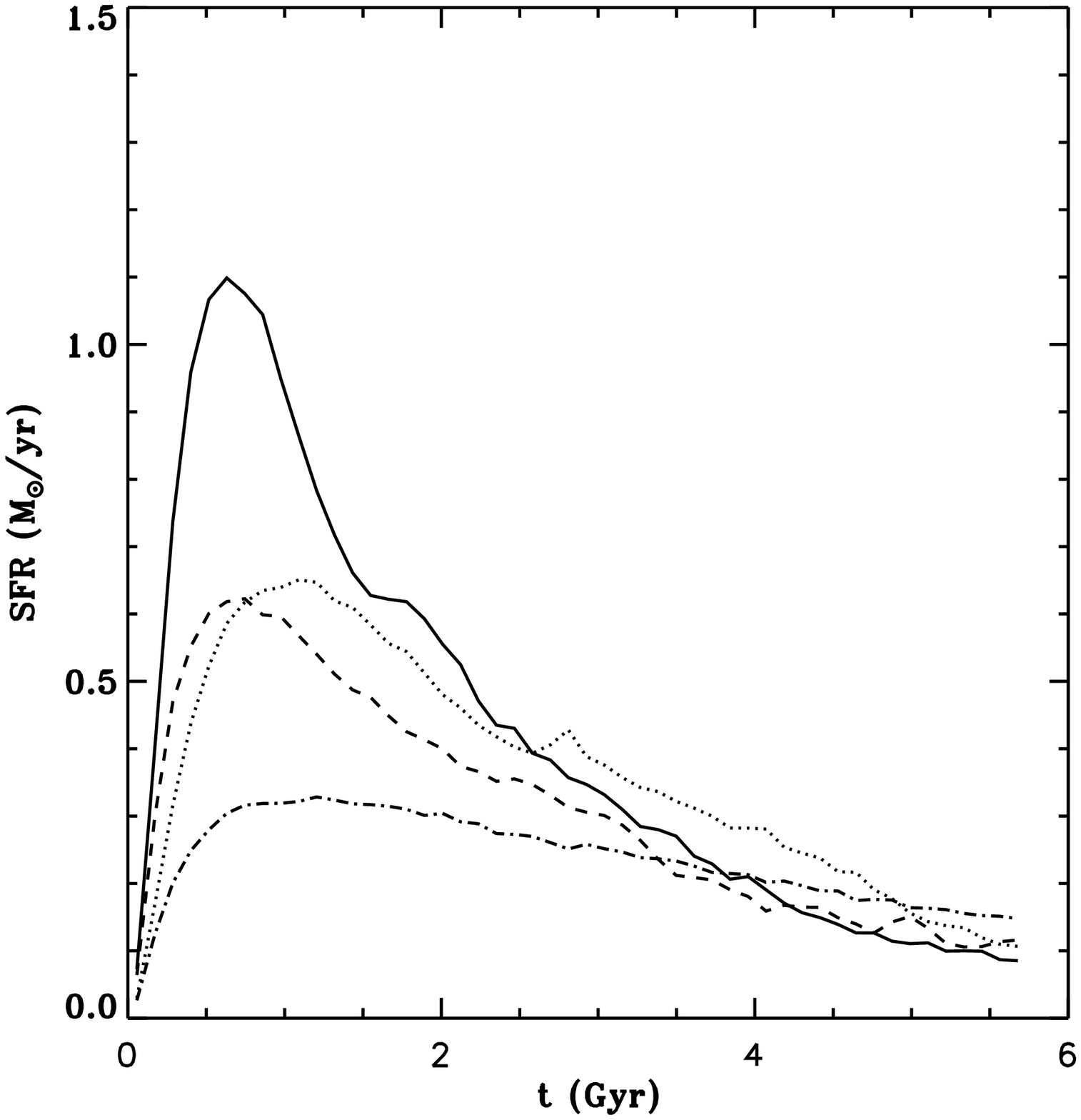}
  \caption{Flux  of mass  (left)  outflowing of  the  $10^{10} \,  \rm
  M_{\odot}$ halos calculated  between $r=[5r_s;7r_s]$ and SFR (right)
  for  the   $\lambda=0.04$  and  $t_0=3\,  \rm   Gyr$  (solid  line),
  $\lambda=0.04$  and $t_0=8\, \rm  Gyr$ (dotted  line), $\lambda=0.1$
  and $t_0=3\,  \rm Gyr$ (dashed line) and  $\lambda=0.1$ and $t_0=8\,
  \rm Gyr$ (dash-dotted line).}
  \label{figure_2}
\end{figure}

Our  various simulations draw  a similar  qualitative picture:  a cold
centrifugally supported disc  form at the halo center  from the inside
out, whose  size depends mainly on  the halo spin  parameter. The star
formation rate in the galaxy rises sharply (see Figure~\ref{figure_2})
and  hot supernovae  bubbles start  to break  out of  the  disc. After
roughly 1  to 3~Gyr, either a galactic wind  develops (for the
low  mass halo)  or a  galactic fountain  sets in  (for the  high mass
halo).   Figure~\ref{figure_1}  shows a  map  of  the gas  temperature
obtained after  6~Gyr in both  cases.  One clearly sees  a noozle-like
structure escaping from a small,  thick disc in the $10^{10}$ M$\odot$
case, while only hot plumes and  cold clumps are seen in the $10^{11}$
M$\odot$ case,  oscillating close  to a large  thin disc, as  they are
confined by a hot, ram pressure driven atmosphere.

In figure~\ref{figure_2} we have plotted  the mass flux flowing out of
the disc,  computed in a  spherical shell of size  $r=[5r_s;7r_s]$, in
the  $10^{10}\,  \rm  M_{\odot}$  case  only, for  various  values  of
$\lambda$ and $t_0$.  The wind  break-out epoch can be determined with
great accuracy:  it appears as a  sharp rise in the  mass outflow rate
curve. Comparing the apparition time of the wind to the SFR, we can asses that a strong SFR at early epoch of galaxy formation creates a young wind that carries out a big quantity of the hot gas of the halo during the first blow.
  Using our  toy  model, we  can  determine the  corresponding
hydrodynamical   efficiency,  which,   depending  on   the  simulation
parameters, varies between  0.8 to 2\%. Injecting these  values in the
toy model  for a  $10^{11}\, \rm M_{\odot}$  halo, we predict  that no
wind can break-out in this  case. Our numerical simulations do confirm
this, as  they show only a  galactic fountain with no  gas leaving the
halo potential well.

When the galactic wind is  fully developped, it shows a typical nozzle
shape, and the mass outflow  rate reaches its asymptotic value, around
0.01 M$_\odot$/yr.  One can  then compute the wind efficiency, defined
as  $\eta_w=\dot M_w/\dot  M_*$, which  ranges from  10 to  20\%, an
order of  magnitude below what  is expected from Lyman  Break galaxies
observations (Martin 1999).

\begin{figure}[h]
  \centering
  \includegraphics[width=6cm]{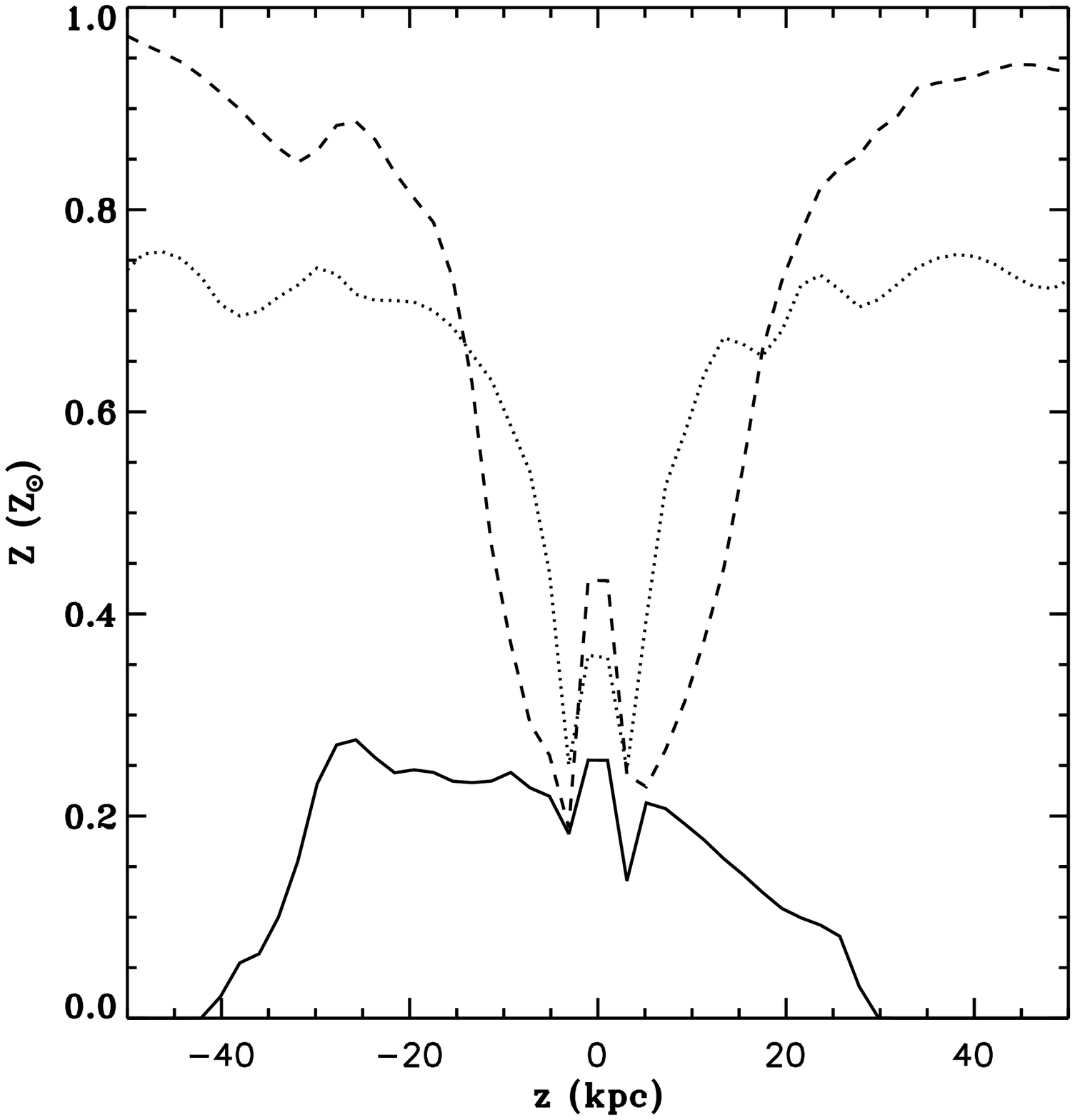}
  \centering
  \includegraphics[width=6cm]{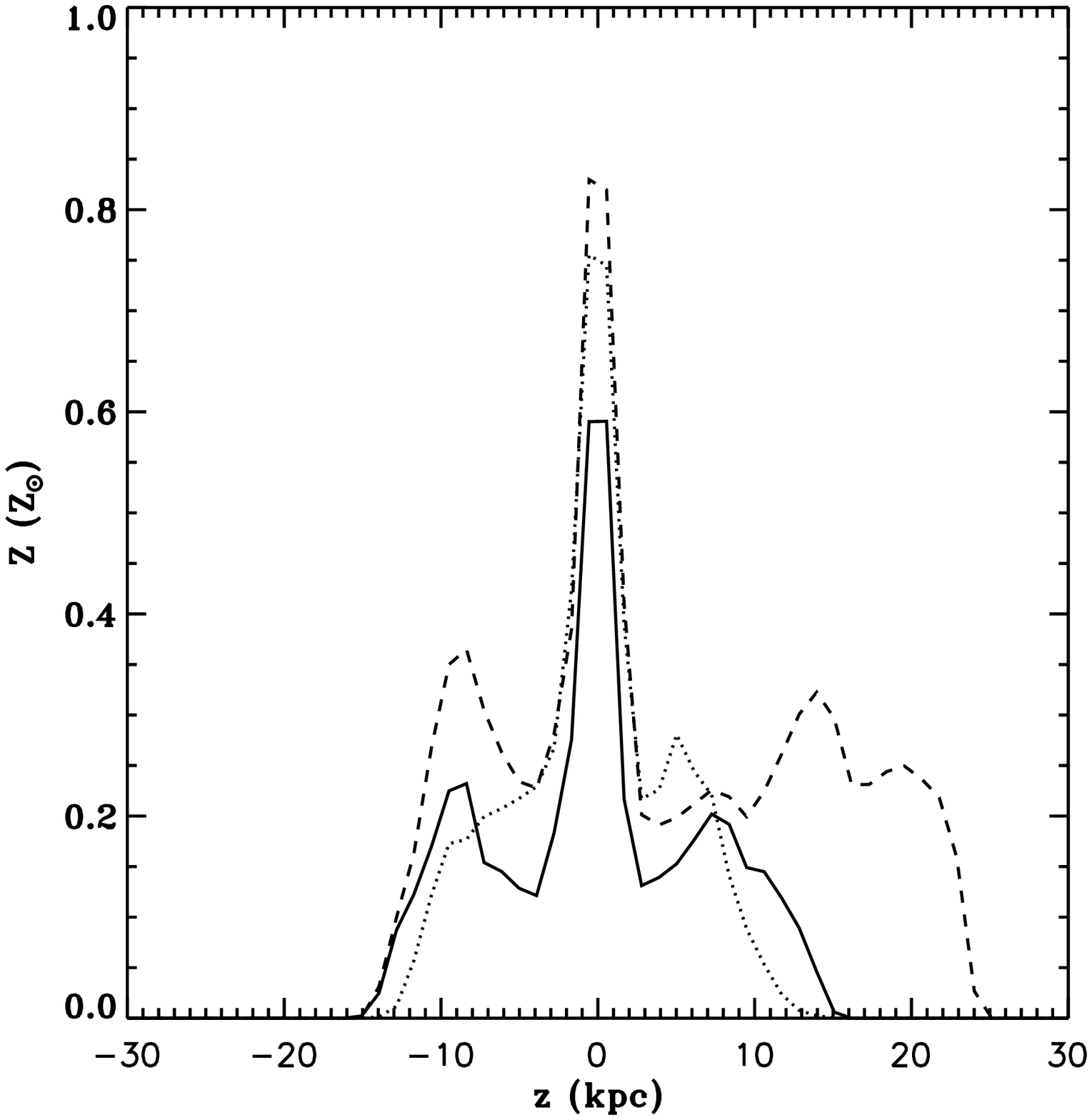}
  \caption{Mean metallicity of the gas as a function of the projected height for the $10^{10}\, \rm  M_{\odot}$ halo (left) and the  $10^{11}\, \rm  M_{\odot}$ halo (right) at different times: $t=3\, \rm Gyr$ (solid line), $t=4.5\, \rm Gyr$ (dotted line) and $t=6\, \rm Gyr$ (dashed line) for $\lambda = 0.1$ and $t_*=8 \, \rm Gyr$.}
  \label{figure_3}
\end{figure}

Figure~\ref{figure_3} shows the mean metallicity projected along the rotation axis of the disc for different times of the galaxy evolution (3 Gyr, 4.5 Gyr, 6 Gyr). The $10^{10}\, \rm  M_{\odot}$ halo begins with a moderately enriched wind ($Z<0.3 \, \rm Z_{\odot}$). The wind propagates in the IGM  with a metallicity growing up to about $\rm Z_{\odot}$ at 6 Gyr, while the disc barely reaches $0.4 \, \rm Z_{\odot}$. The behavior of the $10^{11}\, \rm  M_{\odot}$ halo is quite different because the gas flowing out of the disc falls back. There are two consequences: the gas in the galactic fountain remains at a constant metallicity around $0.2 \, \rm Z_{\odot}$ and the disc progressively accumulates metals up to $\rm Z_{\odot}$. The galactic disc can be seen as a closed system where no gas is lost in the galactic wind.

\section{Conclusion}

Using  a  quiescent model  of  star  formation  in isolated  galaxies,
self-consistently simulated  from a cooling NFW halo,  we have studied
the conditions  for a  galactic wind to  break-out of the  dark matter
halo potential well. Our simulations  have shown that no wind can form
in halo of  mass greater than $10^{11}\, \rm  M_{\odot}$, even for our
most favorable  couple of  halo parameters ($\lambda=0.04$,  $t_0=3 \,
\rm Gyr$). Using a simple toy model, we understand this failure as due
to  the ram pressure  of infalling  material confining  the outflowing
wind.   Using a more  realistic cosmological  setting may  result into
non--spherical  accretion flows,  and  therefore to  a less  stringent
criterion for  a wind  to break-out. A  proper modelling  of starburst
(yet to be invented) might also provide an easier route for increasing
the feedback efficiency of supernovae-driven outflows.

\end{document}